\documentclass[aps,pra,twocolumn,groupedaddress]{revtex4}
\usepackage{graphicx}

\begin{document}

% Use the \preprint command to place your local institutional report
% number in the upper righthand corner of the title page in preprint mode.
% Multiple \preprint commands are allowed.
% Use the 'preprintnumbers' class option to override journal defaults
% to display numbers if necessary
%\preprint{}

%Title of paper
\title{Breakdown of Hydrodynamics in the Radial Breathing Mode\\ of a Strongly-Interacting Fermi Gas}

% repeat the \author .. \affiliation  etc. as needed
% \email, \thanks, \homepage, \altaffiliation all apply to the current
% author. Explanatory text should go in the []'s, actual e-mail
% address or url should go in the {}'s for \email and \homepage.
% Please use the appropriate macro foreach each type of information

% \affiliation command applies to all authors since the last
% \affiliation command. The \affiliation command should follow the
% other information
% \affiliation can be followed by \email, \homepage, \thanks as well.
\author{J. Kinast, A. Turlapov, and J. E. Thomas}
\email[jet@phy.duke.edu]{}
%\homepage[]{Your web page}
%\thanks{}
%\altaffiliation{}
\affiliation{Duke University, Department of Physics, Durham, North
Carolina, 27708, USA}

%Collaboration name if desired (requires use of superscriptaddress
%option in \documentclass). \noaffiliation is required (may also be
%used with the \author command).
%\collaboration can be followed by \email, \homepage, \thanks as well.
%\collaboration{}
%\noaffiliation

\date{\today}

\begin{abstract}
We measure the magnetic field dependence of the frequency and
damping time for the radial breathing mode of an optically
trapped, Fermi gas of $^6$Li atoms near a Feshbach resonance. The
measurements address the apparent discrepancy between the results
of Kinast et al., [Phys.~Rev.~Lett. {\bf 92}, 150402 (2004)] and
those of Bartenstein et al., [Phys.~Rev.~Lett. {\bf 92}, 203201
(2004)]. Over the range of magnetic field from 770 G to 910 G, the
measurements confirm the results of Kinast et al. Close to
resonance, the measured frequencies are in excellent agreement
with predictions for a unitary hydrodynamic gas. At a field of 925
G, the measured frequency begins to decrease  below predictions.
For fields near 1080 G, we observe a breakdown of hydrodynamic
behavior, which is manifested by a sharp increase in frequency and
damping rate. The observed breakdown is in qualitative agreement
with the sharp transition observed by Bartenstein et al., at 910
G.
\end{abstract}

% insert suggested PACS numbers in braces on next line
% insert suggested keywords - APS authors don't need to do this
%\keywords{}

\maketitle
% must follow title, authors, abstract, \pacs, and
%\keywords \maketitle

% body of paper here - Use proper section commands
% References should be done using the \cite, \ref, and \label commands
%\section{}
% Put \label in argument of \section for cross-referencing
%\section{\label{}}
%\subsection{}
%\subsubsection{}

Optically trapped, strongly-interacting Fermi gases of
atoms~\cite{AmScientist,OHaraScience} are possibly the most
convenient and flexible systems for exploring the rich physics of
the BEC-BCS crossover~\cite{Levin}, i.e., the transitional regime
between Bose and Fermi statistics. Atomic systems are unique in
that the state of the gas can be tuned throughout the whole
crossover region, simply  by varying a magnetic field. The last
two years have seen explosive progress in experiments on
degenerate, strongly-interacting Fermi gases which include
anisotropic expansion~\cite{OHaraScience}, studies of universal
interactions~\cite{OHaraScience,SalomonBEC,Grimmbeta}, and
molecular condensates on the BEC side of the
crossover~\cite{GrimmBEC,JinBEC,SalomonBEC,HuletBEC}. Recently,
microscopic evidence for superfluidity has been obtained by
observing preformed pairs on the BCS
side~\cite{Jincondpairs,Ketterlecondpairs} and by measurement of
the gap~\cite{GrimmGap,JinGap}. The focus of this Rapid
Communication is on extension and verification of experiments
which probe the macroscopic properties of the trapped gas by
studying the breathing mode~\cite{Kinast,Bartenstein}. Mapping the
frequency of the mode throughout the BEC-BCS crossover, as a
function of the magnetic field, tests the predictions for the
equation of state and can be used to verify current many-body
calculational methods.

Kinast et al.~\cite{Kinast}, have measured the frequencies and
damping times of the radial breathing mode in an optically trapped
gas of $^6$Li near a Feshbach resonance~\cite{Kinast}. They find
that for magnetic fields in the range of 770-910 G, the
frequencies are close to the hydrodynamic
predictions~\cite{Stringariosc,Heiselbergosc,Hu,Zubarevosc1,Zubarevosc2,Manini}
and that the damping rate drops rapidly as the temperature is
lowered. The temperature dependence of the damping rate  is
consistent with a transition to a superfluid state at low
temperature and inconsistent with a picture of a collisional
normal fluid~\cite{Vichi,Guery,BruunViscous}. Bartenstein et al.,
also have measured the magnetic field dependence of the
frequencies for both the axial and  radial breathing modes of an
optically trapped gas of $^6$Li~\cite{Bartenstein}. For the axial
mode, the agreement with predictions for a hydrodynamic gas is
quite good~\cite{Hu,Manini,Combescot} and the minimum damping rate
is very low. In contrast, the measured frequencies for the radial
mode are 9\% below the predictions of hydrodynamics, and both the
frequency of the radial breathing mode and the damping rate
exhibit an abrupt increase at a magnetic field near 910 G.
Further, the damping rate measured near 910 G by Bartenstein et
al., exceeds the maximum damping rate allowed in a collisional gas
by more than a factor of 5, signaling a possible transition
between a superfluid state and a normal gas.

The discrepancies between the two groups in the  measured
frequencies and damping rates of the radial breathing mode
motivated an additional study of the magnetic field dependence. In
this paper, we describe new measurements of the magnetic field
dependence of the radial breathing mode in an optically trapped,
resonantly interacting gas of $^6$Li, over a magnetic field range
from 750-1114 G (the widest range accessible to us at present).
For fields below 950 G, we find that the measurements confirm the
results obtained by Kinast et al.,~\cite{Kinast}. For fields near
1080 G, we observe a breakdown of hydrodynamic behavior, which is
manifested in a sharp increase in frequency and damping rate. The
observed breakdown is in qualitative agreement with the sharp
transition observed by Bartenstein et al.,~\cite{Bartenstein},
which occurred at  910 G.

In the measurements, we  prepare a highly degenerate 50-50 mixture
of the two lowest spin states of $^6$Li atoms in an ultrastable
CO$_2$ laser trap~\cite{OHaraStable}, using forced evaporation
near a Feshbach resonance~\cite{OHaraScience}. The trap depth is
lowered by a factor of $\simeq 580$ over 4 s, then recompressed to
4.6\% of the full trap depth in 1 s and held for 0.5 s to assure
equilibrium. The number of atoms is determined from the column
density obtained  by absorption imaging on a two-level
state-selective cycling transition~\cite{OHaraScience,Kinast}. In
the measurements of the cloud column density, we take optical
saturation into account exactly and arrange to have very small
optical pumping out of the two-level system. The column density is
integrated in the axial direction and divided by the total atom
number to obtain the one dimensional density $n(x)$, which is
normalized to 1. Fitting the measured distributions of the
expanded cloud with a Thomas-Fermi distribution yields the cloud
radius and the reduced temperature $(T/T_F)_{fit}\simeq 0.1$,
where $T_F$ is the Fermi temperature.

To excite the transverse breathing mode, the trap is turned off
abruptly ($\leq 1\,\mu$s) and turned back on after a delay  of
$t_0=25\,\mu$s. Then the sample is held for a variable time
$t_{hold}$. Finally, the trap is extinguished suddenly, releasing
the gas which is imaged after 1 ms. The temperature increase from
the excitation is  esitmated as $\Delta T/T_F\leq 0.015$ when the
gas thermalizes~\cite{Kinast}.

To determine the breathing mode frequency and damping time, the
transverse spatial distribution of the cloud is fit with a one
dimensional zero-temperature Thomas-Fermi (T-F) profile to extract
the radius, $\sigma_{TF}$ as a function of $t_{hold}$. To acquire
each curve, 60-90 equally spaced values of $t_{hold}$ are chosen
in the time range of interest. The chosen values of $t_{hold}$ are
randomly ordered during data acquisition to avoid systematic
error. Three full sequences are obtained and averaged. The
averaged data is fit with a damped sinusoid
$x_{0}+A\exp(-t_{hold}/\tau)\sin(\omega\, t_{hold} +\varphi )$.
For a $t_0=25 \,\mu$s excitation time, we observe damping times up
to $\tau =7$ ms for both the interacting and noninteracting gas at
4.6\% of full trap depth.

We normalize the  breathing mode frequencies to the transverse
oscillation frequency $\omega_\perp$ for the noninteracting gas at
526 G, where the scattering length is nearly
zero~\cite{zerocross,Grimmzero}. Parametric resonance measurements
yield the trap oscillation frequencies (uncorrected for
anharmonicity) $\omega_x=2\pi\times\,1600(10)$ Hz,
$\omega_y=2\pi\times\,1500(10)$ Hz and
$\omega_z=2\pi\times\,70(5)$ Hz at 4.6\% of full trap depth. As in
our previous study, the frequency $\omega_x$ is also calibrated by
exciting the breathing mode in a noninteracting sample, which
yields  agreement to $\leq 1$\%. In addition, the measured
breathing mode frequency $2\,\omega_x$ is used to scale the value
of $\omega_\perp$ for measurements made on different days, since
small changes in the laser power alter the trap oscillation
frequencies. Typically, we obtain a total atom number
$N=2.0(0.2)\times 10^5$ at temperatures $(T/T_F)_{fit}\simeq 0.1$.
From the measured trap frequencies, we find (after correction for
anharmonicity using Eq.~\ref{eq:nuperp} below)
$\omega_\perp=\sqrt{\omega_x\omega_y} = 2\pi\times\,1596 (7)$ Hz,
and
$\bar{\omega}=(\omega_x\omega_y\omega_z)^{1/3}=2\pi\times\,560(13)$
Hz. For these parameters, the typical Fermi temperature for a
noninteracting gas is $T_F=(3N)^{1/3}\hbar\bar{\omega}/k_B\simeq
2.2\,\mu$K, small compared to the final trap depth of $35\,\mu$K.

Anharmonicity arising from the gaussian trapping potential reduces
the measured frequencies $\omega^{\text{meas}}$ from their
zero-energy harmonic oscillator values $\omega$.  The transverse
oscillation frequency $\omega_\perp$ for the noninteracting gas at
526 G is corrected using the formula~\cite{Stringarishift},
\begin{equation}
\omega_\perp/\omega_\perp^{\text{meas}}=1\,
+\,(6/5)\,M\omega_\perp^2x_{\text{Brms}}^2/(b_B^2U) ,
\label{eq:nuperp}
\end{equation}
where $U$ is the trap depth, $M$ is the $^6$Li mass,  and
$b_B=\sqrt{1+(\omega_x t)^2}=10.3$ is the ballistic expansion
factor for the release time $t=1$ ms. Note that the rms width for
the ballistically expanding gas is
$x_{\text{Brms}}=\sigma^B_{TF}/\sqrt{8}$, where $\sigma^B_{TF}$ is
the transverse T-F radius of the ballistically expanded cloud at
526 G.

The measured hydrodynamic frequencies $\omega_H$ are corrected for
anharmonicity using~\cite{Stringarishift}
\begin{equation}
\omega_H/\omega_H^{\text{meas}}=1\, +\,
(32/25)\sqrt{10/3}\,M\omega_\perp^2x_{\text{rms}}^2/(b_H^2U).
\label{eq:nuhydro}
\end{equation}
 Here, $b_H$ is the hydrodynamic expansion factor~\cite{OHaraScience,Menotti}, 11.3 after 1 ms, and $x_{\text{rms}}=\sigma_{TF}/\sqrt{8}$
 is the rms width for the interacting gas after expansion. Note that $b_H$ rather than $b_B$ is used
 to estimate the trapped cloud rms radius because we
 observe anisotropic expansion of the gas~\cite{OHaraScience} which is in good
 agreement with the predictions of
 hydrodynamics~\cite{OHaraScience,Menotti}.

\begin{figure}[htb]
\includegraphics[width=3.3in]{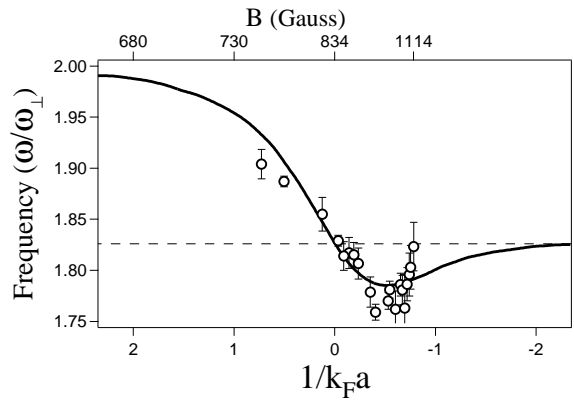}
\caption{Magnetic field dependence of the frequency $\omega$ of
the radial breathing mode. Solid line is the  theory based on
superfluid hydrodynamics from Hu et al.~\cite{Hu}. Dashed line is
the hydrodynamic frequency for a unitarity gas, $\sqrt{10/3}$, as
predicted at resonance. Note that the top (magnetic field) axis is
not linear. \label{fig:magdepfreq}}
\end{figure}

The extracted ratios of the corrected frequencies,
$\omega_H/\omega_\perp$, and the measured damping ratios
$1/(\omega_\perp\tau)$  are given in Table 1 as a function of
magnetic field. The frequencies  are plotted in
Fig.~\ref{fig:magdepfreq} as a function of magnetic field.  The
lower horizontal scale gives $1/(k_Fa)$, where $k_F$ is the Fermi
wavevector at the trap center $\hbar k_F=\sqrt{2M\,k_BT_F}$ and
$a=a(B)$ is the scattering length. Note that we assume that the
Feshbach resonance is located at 834 G, consistent with the best
recent estimates~\cite{SchunckFeshbach,BartensteinFeshbach}.
The solid curves in Fig.~\ref{fig:magdepfreq} show the predictions
for the frequency by Hu et al. based on assumptions of superfluid
hydrodynamics~\cite{Hu,Hudetails}.

\begin{figure}[htb]
\includegraphics[width=3.3in]{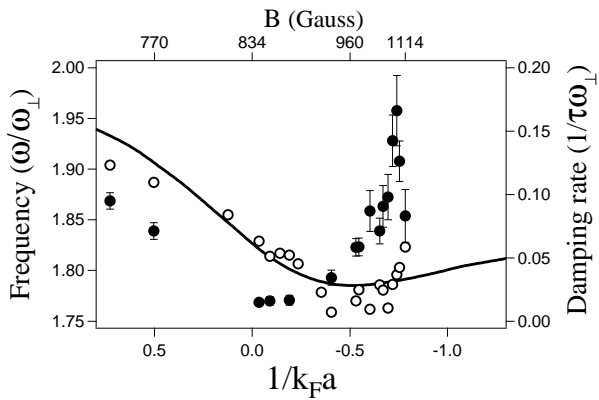}
\caption{Magnetic field dependence of the damping rate $1/\tau$ of
the radial breathing mode. Damping rate (solid circles, axis
labelled on the right); frequency (data--open circles, theory--
solid curve, axis labelled on the left). Solid curve (predicted
frequency). The maximum damping ratio occurs at 1080G and exceeds
the maximum 0.09 permitted for collisional dynamics by a factor of
two. Note that the top (magnetic field) axis is not linear.
\label{fig:magdepgamma}}
\end{figure}

The data exhibit several interesting features. For fields near
resonance, the measured radial breathing mode frequencies are in
very good agreement with hydrodynamic
theory~\cite{Stringariosc,Heiselbergosc,Hu,Zubarevosc1,Zubarevosc2},
confirming our previous measurements~\cite{Kinast}. Further, the
measured breathing mode frequency at 840 G,
$\omega_H/\omega_\perp=1.829 (.006)$, is very close to that
predicted for a unitary-limited, hydrodynamic Fermi gas, where
$\omega_H/\omega_\perp=\sqrt{10/3}=1.826$.
Fig.~\ref{fig:magdepgamma} shows  that the damping rates are small
near resonance. The maximum damping time of 7 ms, obtained near
resonance, corresponds to a minimum damping ratio
$1/(\omega_\perp\tau)=0.014$, or about 20 periods of hydrodynamic
oscillation.

Away from resonance, the data in Fig.~\ref{fig:magdepfreq} exhibit
qualitative agreement with the hydrodynamic theory of Hu et
al.~\cite{Hu}. However, the frequencies measured well below
resonance, at 750 G and 770 G,  are lower than predicted by
approximately two standard deviations. Further, above resonance,
the measured magnetic field dependence of the frequency is
compressed. In addition, there is a region where the data are
consistently below the theory and this is followed by a region
where the data are consistently above. The frequency reaches its
lowest value at 925 G, dropping below the theory by three standard
deviations, and then it rises rapidly at 1080 G. At the highest
field achievable with our magnets, 1114 G,  the frequency is well
above the theoretical prediction, but does not reach the
asymptote, $2\,\omega_x$, for a noninteracting gas. In this region
above resonance, we observe a clear breakdown of hydrodynamics,
both in the rapid change in frequency at 1080 G and in the damping
rate. A transition is obvious in Fig.~\ref{fig:magdepgamma}, which
shows the damping rate  as a function of $B$. Near 1080 G, the
damping rate increases to $0.17\,\omega_\perp$, well above the
maximum $0.09\,\omega_\perp$ attainable by collisional
hydrodynamics~\cite{Vichi,Guery}.  The observed behavior is
therefore in qualitative agreement with the abrupt frequency
change and increased damping observed by the Innsbruck
group~\cite{Bartenstein}. However, our results differ from those
of Bartenstein et al., in two important respects: i) Near
resonance, our observations for the radial breathing mode agree
with hydrodynamic theory; ii) the field of 1080 G, at which we
observe the complete breakdown of hydrodynamics and rapid damping,
is substantially higher than the 910 G value observed by the
Innsbruck group~\cite{Bartenstein}.

The Innsbruck group has suggested that the breakdown may arise
when the zero temperature BCS energy gap $\Delta$ is smaller than
the collective mode energy $\hbar\omega$~\cite{Bartenstein}.
Zubarev~\cite{Zubarevosc1,Zubarevosc2} has determined the trap
averaged value of $\Delta$ and shows that if the abrupt transition
occurs at 910 G for the Innsbruck group, then for the conditions
of our trap, we should observe the transition near 1000 G,
assuming $3\times 10^5$ atoms. Using similar estimates, we find
that at 910 G, $\Delta\simeq 2\hbar\omega$ for the conditions of
the Innsbruck group and $\Delta\simeq\hbar\omega$ for our trap
conditions at 1080 G. Since single particle excitations have a
minimum energy of $2\Delta$, perhaps the condition $\Delta
=\hbar\omega$ underestimates the required collective mode energy
for complete pair breaking. However, it has been noted by
Heiselberg~\cite{Heiselbergpairing} that in-gap surface modes have
a smaller energy than $\Delta$, but scale similarly, possibly
accounting for the observed breakdown at $\Delta > \hbar\omega$.

For the conditions of the Innsbruck group, the breakpoint occurs
at the same magnetic field of 910 G at a shallow trap depth and
for a trap depth increased by a factor of 9~\cite{Bartenstein}.
This magnetic field independence was first explained by the
Innsbruck group and arises as follows. In their case (but not in
ours) the axial confinement is provided primarily by the bias
field magnets, and the ratio of the Fermi energy to
$\hbar\omega_\perp$ and hence to $\hbar\omega$ {\it decreases}
with increasing trap depth. Then, using the most recent values of
the scattering length~\cite{BartensteinFeshbach}, we find the
increase in $k_F$ in the exponent of the pairing gap compensates
for this reduction at 930 G, consistent with their argument.

 We can obtain
an alternative estimate of the magnetic field at which an abrupt
change occurs using an idea due to Falco and
Stoof~\cite{StoofCrossover}. They suggest that when the Feshbach
($v=38$) molecular state in the singlet potential has a higher
energy than the two-particle Fermi energy (relative to the zero of
the triplet potential), the system becomes BCS-like. In this case,
the system may not be interacting strongly enough to remain
superfluid at the temperatures we achieve. A simple estimate of
the breakdown magnetic field is then obtained from
\begin{equation}
\frac{\hbar^2}{Ma^2}=2\langle \epsilon_F(\mathbf{x})\rangle .
\label{eq:moleculenergy}
\end{equation}
We assume that near resonance, interactions make the same
contribution to the chemical potential of  free triplet atoms and
to very weakly bound atoms in large singlet molecules. We also
assume the energy of the molecular state is
$\epsilon_b=\hbar^2/Ma^2$ relative to the zero of energy in the
triplet potential, which is valid in the two-body case near the
Feshbach resonance. Using the trap averaged local Fermi energy, we
obtain $2\langle \epsilon_F(\mathbf{x})\rangle= (5/4)\,
k_BT_F=(5/8)\, \hbar^2 k_F^2/M$, so that
$1/(k_Fa)=-\sqrt{5/8}=-0.79$. From Table~\ref{table:1}, we see
that at $B=1080$ G, $1/(k_F a)=-0.74$, in  reasonable agreement.
Unfortunately, our simple hypothesis cannot explain the
insensitivity of the B field value with respect to trap depth
observed by Bartenstein et al.: Scaling our Fermi energy down from
$2.2\,\mu$K to $1.2\,\mu$K for the conditions of the Innsbruck
group, we predict a transition near $B=970$ G, higher than
observed.

\begin{table}
\centering
\begin{tabular}{|c|c|c|c|}
  \hline
  % after \\: \hline or \cline{col1-col2} \cline{col3-col4} ...
  B(G) & $1/(k_Fa)$ & $(\omega/\omega_\perp)$& $1/(\omega_\perp\tau)$  \\
  \hline
  750  & 0.728 &1.904(.015)&0.095(.009)\\ \hline
  770  & 0.504 &1.887(.005)&0.071(.007) \\ \hline
  815  & 0.124 &1.855(.017)&  *           \\ \hline
  840  &-0.034 &1.829(.006)&0.014(.001) \\ \hline
  850  &-0.090 &1.816(.014)&0.016(.005) \\ \hline
  860  &-0.142 &1.817(.015)&   *         \\ \hline
  870  &-0.190 &1.815(.012)&0.017(.004) \\ \hline
  880  &-0.235 &1.807(.015)&    *        \\ \hline
  910  &-0.354 &1.779(.015)&    *        \\ \hline
  925  &-0.405 &1.759(.008)&0.034(.006) \\ \hline
  969  &-0.531 &1.766(.007)&0.058(.007) \\ \hline
  975  &-0.546 &1.778(.007)&0.059(.007) \\ \hline
  1000 &-0.603 &1.763(.018)&0.087(.016) \\ \hline
  1025 &-0.652 &1.786(.011)&0.071(.010) \\ \hline
  1035 &-0.670 &1.781(.015)&0.091(.016) \\ \hline
  1050 &-0.695 &1.764(.033)&0.098(.020) \\ \hline
  1065 &-0.719 &1.786(.016)&0.142(.020) \\ \hline
  1080 &-0.740 &1.796(.021)&0.166(.028) \\ \hline
  1090 &-0.754 &1.803(.021)&0.126(.016) \\ \hline
  1114 &-0.783 &1.823(.024)&0.083(.021) \\ \hline
  \hline
  \end{tabular}
\caption{Breathing mode frequencies $\omega$ and damping rates
$1/\tau$ as a function of applied magnetic field B. Error
estimates include the statistical error from the fit only.
$^*$Denotes data taken with a $50\,\mu$s excitation
time~\cite{Kinast} for which we omit the damping rate. $a(B)$ is
determined from Ref.~\cite{BartensteinFeshbach}}\label{table:1}
\end{table}

In summary, we have measured the magnetic field dependence of the
frequencies and damping rates of the radial breathing mode for a
strongly interacting Fermi gas of $^6$Li. The measurements are in
very good agreement with the predictions of hydrodynamics near the
Feshbach resonance at 834 G,  but we observe a breakdown of
hydrodynamics away from resonance. Beginning at higher fields near
925 G we observe first a frequency decrease, well below
predictions, and then an abrupt frequency increase at a field of
1080 G, accompanied by rapid damping, which exceeds the maximum
damping for a collisional normal fluid.

We are indebted to R.Hulet and R. Grimm for providing preliminary
estimates of the location of the Feshbach resonance and we thank
R. Grimm, P. Julienne, and C. Williams for providing detailed
information on the shape of the Feshbach resonance in advance of
publication.

This research is supported by the  Chemical Sciences, Geosciences
and Biosciences Division of the Office of Basic Energy Sciences,
Office of Science, U. S. Department of Energy, the Physics
Divisions of the Army Research Office  and the National Science
Foundation, and  the Fundamental Physics in Microgravity Research
program of the National Aeronautics and Space Administration.

%\bibliography{MagField}

\begin{thebibliography}{34}
\bibitem[{Str()}]{Stringarishift}
\bibinfo{note}{These frequency shifts were derived by S. Stringari, private
  communication.}
\end{thebibliography}
% Create the reference section using BibTeX:

\end{document}